%%%%%%%%%%%%%%%%%%%%%%%%%%%%%%%%%%%%%%%%%%%%%%%%%%%%%%%%%%%%%%%%%%%%%%%%%%%%%%%
\input harvmac
%\draftmode
\overfullrule=0pt
\abovedisplayskip=12pt plus 3pt minus 3pt
\belowdisplayskip=12pt plus 3pt minus 3pt
\sequentialequations
%macros
%
\def\tilde{\widetilde}
\def\bar{\overline}
\def\to{\rightarrow}

\font\zfont = cmss10 %scaled \magstep1
\font\litfont = cmr6

\def\bigone{\hbox{1\kern -.23em {\rm l}}}
\def\ZZ{\hbox{\zfont Z\kern-.4emZ}}
\def\half{{\litfont {1 \over 2}}}

% References

\def\BECKERS{K. Becker and M. Becker, {\it ``M-Theory on Eight
Manifolds''}, hep-th/9605053, Nucl. Phys. {\bf B477} (1996) 155.}
\def\SVW{S. Sethi, C. Vafa and E. Witten, {\it ``Constraints on
Low-Dimensional String Compactifications''}, hep-th/9606122, 
Nucl. Phys. {\bf B480} (1996) 213.}
\def\VAFWIT{C. Vafa and E. Witten, {\it ``A One Loop Test Of String
Duality''}, hep-th/9505053, Nucl. Phys. {\bf B447} (1995) 261.}
\def\DLM{M.J. Duff, J.T. Liu and R. Minasian, {\it 
``Eleven-dimensional Origin of String/String Duality: a 
One Loop Test''}, hep-th/9506126, Nucl. Phys. {\bf B452} (1995) 261.}
\def\HORWIT{P. Horava and E. Witten, {\it ``Heterotic and Type I 
String Dynamics from Eleven Dimensions''}, hep-th/9510209,
Nucl. Phys. {\bf B460} (1996) 506.}
\def\SEIWIT{N. Seiberg and E. Witten, {\it ``Comments on String
Dynamics in Six Dimensions}, hep-th/9603003, 
Nucl. Phys. {\bf B471} (1996) 121.}
\def\WITFLUX{E. Witten, {\it ``On Flux Quantization in M-theory and
the Effective Action''}, hep-th/9609122.}
\def\BERGSHOEFF{E. Bergshoeff, H.J. Boonstra and T. Ort\'in, {\it
``S-duality and Dyonic p-brane Solutions in Type II String Theory''},
hep-th/9508091.}
\def\JOYCE{D. Joyce, {\it ``Compact Riemannian Manifolds with Holonomy
spin(7)''}.}
\def\BORCEA{C. Borcea, {\it ``K3 Surfaces with Involution and 
Mirror Pairs of Calabi-Yau Manifolds''}, to appear in {\it ``Essays on
Mirror Manifolds (II)''}.}
\def\SHIBAJI{S. Roy, {\it ``An Orbifold and Orientifold of Type IIB
Theory on K3 $\times$ K3''}, hep-th/9607157.}
\def\SENORB{A. Sen. {\it ``Orbifolds of M Theory and String Theory''},
hep-th/9603113, Mod. Phys. Lett. {\bf A11} (1996) 1339.}
\def\DM{K. Dasgupta and S. Mukhi, {\it ``Orbifolds of M-theory''},
hep-th/9512196, Nucl. Phys. {\bf B465} (1996) 399.} 
\def\ALOK{A. Kumar and K. Ray, {\it ``Compactifications
of M-Theory to Two Dimensions''}, hep-th/9604164,
Phys. Lett. {\bf B383} (1996) 160.} 
\def\ORI{O. Ganor, {\it ``Compactification of Tensionless 
String Theories''}, hep-th/9607092.}

{\nopagenumbers
\Title{\vtop{\hbox{hep-th/9612188}
\hbox{TIFR/TH/96-61}}}
{\centerline{A Note on Low-Dimensional String Compactifications}}
\centerline{Keshav Dasgupta\foot{E-mail: keshav@theory.tifr.res.in}
and Sunil Mukhi\foot{E-mail: mukhi@theory.tifr.res.in}}
\vskip 4pt
\centerline{\it Tata Institute of Fundamental Research,}
\centerline{\it Homi Bhabha Rd, Bombay 400 005, 
India}
\ \medskip
\centerline{ABSTRACT}

We study supersymmetric compactifications of type II strings on
eightfolds to two dimensions. It is demonstrated that the type IIB
string on an eightfold is free of gravitational anomalies. T-duality
requires that this theory when further compactified on a circle must
have a vacuum momentum; this is explicitly shown to be present and to
have the right value. A subtlety in the relation of IIB
compactifications and M-theory orientifolds to two dimensions is
pointed out.

\ \vfill 
\leftline{December 1996}
\eject} 
\ftno=0
\newsec{Introduction} 

Compactifications of string theory to 2 spacetime dimensions are
interesting for various reasons, one of which is the relationship to M
and F theory in 3 and 4 dimensions respectively (this holds for the
type IIA string). Certain consistency conditions arise when we require
the compactification to satisfy the equations of
motion\ref\beckers{\BECKERS}\ref\svw{\SVW}.  The type IIB string on an
8-fold leads to chiral theories in two dimensions. Since type IIA and
IIB are T-dual to each other after compactification on a circle, there
are various interconnections between the above properties. This will
be the subject of the present paper.

\newsec{Type IIA on eightfolds}

Compactification of the type IIA string on an 8-fold is potentially
destabilised by a term of the form $\int B\wedge I_8$ which arises at
1-loop level\ref\vafwit{\VAFWIT}. Here $B$ is the 2-form gauge field, and
$I_8$ is a linear combination of the Pontryagin classes $p_2$ and
$p_1^2$, whose integral on an eightfold gives the Euler
characteristic $\chi$. (A similar term $\int C\wedge I_8$ arises in
M-theory\ref\dlm{\DLM}, where this time $C$ is the M-theory
3-form field.) A naive compactification of type IIA on an eightfold
thus gives a tadpole term $\int B$ in two dimensions (and analogously,
M-theory gives a tadpole $\int C$ in three dimensions), proportional
to the value of $\chi$.

It was observed in Ref.\beckers\ that for M-theory compactifications,
this tadpole also receives contributions from the classical term $\int
C\wedge dC \wedge dC$ if we allow a background value of the 3-form
field on the eightfold. This contribution is proportional to $\int
dC\wedge dC$ over the eightfold. Clearly the analogous result holds
also for type IIA compactifications. We will discuss signs and factors
presently. 

Finally, in Ref.\svw\ it was noted that the presence of branes in the
vacuum, filling spacetime, contributes to the tadpole as well. (Thus
we need 1-branes for IIA and 2-branes for M-theory). Each brane
modifies the tadpole by an integer value.

Combining the above results, the condition for a consistent type IIA
compactification on an eightfold are:
\eqn\iiacond{
{\chi\over 24} - {1\over 8\pi^2} \int dC\wedge dC - n =0}
where $n$ is the number of branes\foot{Note that we differ from
Ref.\beckers\ in the sign of the second term above. We are grateful to
E. Witten, K. Becker and M. Becker for correspondence on this 
point.}.

To check the signs and factors in this expression, let us compare type
IIA on $K3\times K3$ with the heterotic string on $T^4\times
K3$. These two theories are dual to each other, from six-dimensional
string-string duality. For the heterotic string on $K3$, we have a
condition relating the instanton number of the background gauge fields
on $K3$ and the number of M-theory 5-branes in the vacuum, which can
arise if we view the heterotic string as M-theory on
$S^1/Z_2$\ref\horwit{\HORWIT}. 
This condition may be written\ref\seiwit{\SEIWIT}:
\eqn\hetcond{
24 - {1\over 16\pi^2} \int \tr F\wedge F - n = 0} 
where the second term is integrated over $K3$. n is the number of
M-theory 5-branes, but in the present context they are wrapped on
$T^4$ to become 1-branes. Clearly the number of 1-branes is bounded
above by 24. Moreover, the sign of the second term (the instanton
number) is such that instantons, rather than anti-instantons,
contribute a negative integer to the equation.

Equation \iiacond\ integrated over $K3\times K3$ gives precisely this
result, since the Euler of $K3 \times K3$ is $(24)^2$, while the
second term gives rise (generically) to $U(1)$ gauge fields on one
$K3$ and their Pontryagin class is then integrated over the second
$K3$. Note that this term is integer in Eq.\iiacond\ as long as
$[{dC\over 2\pi}]$ is an integral cohomology class, which (as shown in
Ref.\ref\witflux{\WITFLUX})
is true whenever $\chi\over 24$ is integral. Finally,
the last term is the number of IIA 1-branes, which are M-theory
2-branes wrapped on the 11th dimension. So the last terms are
electric-magnetic dual between the two theories, as expected.

This confirms Eq.\iiacond, and shows that the number of branes in the
vacuum is bounded above by the Euler characteristic of the eightfold
(this would not have been true if the second term had the opposite
sign)\foot{An independent check of the sign in Eq.\iiacond\ can be
made by examining Eq.(4.4) of Ref.\witflux. This gives the same
result.}.

Note that the second term of Eq.\iiacond\ takes values congruent to
$1\over 4$ mod integers if $[{dC\over 2\pi}]$ is a half-integral 
cohomology class. In this case, $\chi\over 24$ must also be congruent
to $1\over 4$ mod integers.

\newsec{Analogous issues for type IIB}

On compactifying type IIB on a circle, it becomes equivalent to IIA on
a circle under T-duality. Thus it must possess the dimensional
reductions of both the classical term $\int B\wedge dC \wedge dC$ and
the one-loop term $\int B \wedge I_8$. After reducing on a circle, the
$B$-field of type IIA becomes a 1-form $A$ which measures the winding
charge with respect to that circle. Under T-duality this, in turn,
becomes the Kaluza-Klein 1-form arising by reduction of the 10D metric
of IIB on the circle. Thus we must look for terms in type IIB which
reduce to $\int A\wedge dC \wedge dC$ and $A\wedge I_8$ in 9
dimensions. 

To find the first of these is reasonably straightforward, except for a
familiar subtlety: for nonzero values of the self-dual 4-form $D^+$,
the type IIB string does not have a covariant action. Yet we need
precisely terms depending on $D^+$, since that field reduces to the
3-form $C$ in 9d. One way out is to make use of the so-called
non-self-dual (NSD) \ref\bergshoeff{\BERGSHOEFF}\ action for type IIB in
10d. In this formulation, one starts by ``forgetting'' the
self-duality condition on $D$. An action can then be written down,
with the property that its equations of motion reduce to the correct
ones after imposition of the self-duality constraint by hand.

In this formalism the $D$ field has a kinetic term $\int dD\wedge
*dD$. This term involves the metric via the operation of taking the
Poincare dual. However, ``morally speaking'' it is topological, since
the 5-form $dD$ is eventually set equal to its dual. Thus on
compactification to 9d, it can give rise to a topological term. The $D$
field reduces to a 3-form in 9d, while one of the index contractions
requires the metric component $g_{\mu 10}$ which is just the KK gauge
field $A_\mu$. (we label the dimensions
$(x^1,x^2,\ldots x^{10})$ where $x^1$ is the time). As a result, one
gets the desired term $\int A\wedge dC \wedge dC$ in 9d. 

The one-loop term is far more subtle. It is known that in 10
dimensions there is no one-loop correction in type IIB analogous to
the term $\int B\wedge I_8$ in type IIA. Moreover, one can easily
convince oneself directly that there is no purely gravitational term
that one can write down in 10d which reduces to $\int A\wedge I_8$ in
9d with $A$ being the KK gauge field. In fact, we will argue that no
modification is required in 10d to the type IIB action, but as soon as
one compactifies on a circle, however large, there is a
radius-dependent term of the desired form in 9d. We will return to
this point below.

\newsec{Type IIB on eightfolds: the anomaly}

Consider the supersymmetric compactifion of type IIB on an
eightfold. We will consider manifolds of holonomy
$spin(7)$, $SU(4)$ and subgroups of $SU(4)$. Joyce 
eightfolds\ref\joyce{\JOYCE}\
of $spin(7)$ holonomy lead to $(0,2)$ spacetime supersymmetry in 2
dimensions, while Calabi-Yau eightfolds of $SU(4)$ holonomy give
$(0,4)$ supersymmetry. Other interesting cases are $K3\times K3$
leading to $(0,8)$ supersymmetry, and $T^8/Z_2$ (an orbifold which
cannot be blown up to a smooth manifold, but gives consistent string
compactifications all the same) where the supersymmetry is
$(0,16)$. In each case, the supersymmetry is counted in terms of
one-component Majorana-Weyl spinors.

The number of scalars from a CY eightfold  in 2 dimensions is counted as follows: the 10d
metric $g_{MN}$ gives rise to $h_{11}+2h_{31}$ non-chiral scalars,
while the two 2-form fields $B_{MN}$ and ${\tilde B}_{MN}$
produce $2h_{11}$ non-chiral scalars. Two more come from the 10
dimensional scalars of type IIB. Finally, the self-dual 4-form 
$D^+_{MNPQ}$ leads to $(b_4^-)=h_{22}^- + 2h_{31}^-$ scalars of 
$+$ chirality and $(b_4^+)= 2 + h_{22}^+ + 2h_{31}^+$ scalars of $-$
chirality. 
 
Combining, we have $n(\phi^+) = 3h_{11} + 2h_{31}+2+b_4^-$ and
$n(\phi^-)= 3h_{11}+2h_{31}+2+b_4^+$. With $(0,N)$ supersymmetry, the
supergravity multiplet is $(g_{\mu\nu}, \phi, N\psi_{\mu}^-,
N\psi^+)$. Let us assume that we have some number $n_+/N$ of chiral
matter multiplets $(N\phi^+, N\psi^+)$ (thus, $n_+$ counts the number
of individual scalars, which is more convenient), along with
$n_-^\phi$ anti-chiral scalars $\phi^-$, and $n_-^\psi$ anti-chiral
Majorana-Weyl fermions $\psi^-$ which are supersymmetry singlets.

Since one scalar goes into the gravity multiplet, we have
$n_+ = n(\phi^+) - 1$, $n_-^{\phi}= n(\phi^-)-1$. The difference
is given by
\eqn\differ{
n_+ - n_-^\phi = b_4^- - b_4^+ = -\tau}
where $\tau$ is the signature of the eightfold. 

The gravitational anomaly from the supergravity multiplet, combining
contributions from the gravitinos and the spin-$\half$ fermions, is
$N$ times the anomaly polynomial. Chiral scalars contribute $1\over
12$ in the same units, and Majorana-Weyl fermions $1\over 24$. Thus
the total anomaly is proportional to
\eqn\totalanomaly{
N + {n_+ - n_-^{\phi}\over 12} + {n_+ - n_-^{\psi}\over 24}}

The fermions are counted as follows: the spin-$3\over 2$
particles in 2 dimensions come from the gravitinos in 10 dimensions,
and their number is given by the Dirac index of the internal manifold.
An equal number of spin-$1\over 2$ particles in two dimensions 
come from the spin-$\half$ fermion in 10 dimensions. All these go into the
supergravity multiplet. Additional spin-$\half$ fermions arise from
the 10d gravitinos, and their 
number is given in terms of the Rarita-Schwinger index. Therefore
\eqn\rarschw{
n_-^{\psi} - n_+ = 2~ind(\Dsl_{3/2})}
where the 2 on the RHS comes from the fact that there are 2 gravitinos
in 10 dimensions.

In terms of Pontryagin classes, we have
\eqn\rspont{
ind(\Dsl_{3/2})= {37\over 720} p_1^2 - {31\over 180}p_2}
where
\eqn\pontclas{
\eqalign{p_1 &= -{1\over 2} tr R^2\cr
p_2 &= -{1\over 4} tr R^4 + {1\over 8}(tr R^2)^2 }}

Using the additional relations
\eqn\addl{
\eqalign{\chi &= {p_2\over 2} - {p_1^2\over 8}\cr
ind(\Dsl_{1/2}) &= {N\over 2} = {1\over 1440}({7\over 4}p_1^2 -
p_2)\cr}}
we find that
\eqn\solvepi{
\eqalign{
p_2 &= 120N + {7\chi\over 3}\cr
p_1^2 &= 480N + {4\chi\over 3}\cr}}

>From these relations, it follows that
\eqn\rsindsolv{
ind(\Dsl_{3/2}) = 4N - {\chi\over 3}}

The anomaly thus becomes
\eqn\anomthus{
N - {\tau\over 12}  - {8N-2\chi/3\over 24}}
and the condition for anomaly cancellation reduces to
\eqn\anomcanc{
24N - 3\tau + \chi=0}

This is actually an identity, as can be easily seen by replacing each
term by its value in terms of Pontryagin classes, using Eq.\addl\ and
the Hirzebruch signature theorem
\eqn\hirz{
\tau = {1\over 45}(7p_2-p_1^2)}

Thus we have demonstrated explicitly that the type IIB string on any
eightfold preserving some supersymmetry is anomaly-free in 2d.

\newsec{Type IIA and IIB on eightfolds: the spectrum}

We can easily write down the complete spectrum of type IIA and IIB on
an eightfold. For type IIA, the supersymmetry is labelled $({N\over
2}, {N\over 2})$ and the supermultiplets are the supergravity
multiplet $(g_{\mu\nu}, \phi, {N\over 2}\psi_{\mu}^+, {N\over
2}\psi_{\mu}^-, {N\over 2}\psi^+, {N\over 2}\psi^-)$.  We assume a
number $(2n_+/N)$ of (non-chiral) matter multiplets $({N\over
2}\phi^+, {N\over 2}\phi^-, {N\over 2}\psi^+, {N\over 2}\psi^-)$. Then
it is easy to see that for a CY
\eqn\iiaspec{
{n_+\over 2} = h_{11} + h_{21} + h_{31} = {\chi\over 6} - 8 + 2h_{21}}
where the last equality follows from the identity\svw:
\eqn\svwidentity{
h_{11} - h_{21} + h_{31} = {\chi\over 6}-8}

For type IIB, we need to use the index calculations of the previous
section in addition to the cohomology of the eightfold. We have seen
at the beginning of the previous section that $n_+$ and $n_-^\phi$,
the number of $+$ and $-$ chirality scalars are given by
\eqn\nplusminus{
\eqalign{n_+ &= 3h_{11} + 2h_{31} + 1 + b_4^- \cr
n_-^\phi &= 3h_{11} + 2h_{31} + 1 + b_4^+ \cr}}

Following Eq.\svwidentity, this can be rewritten
\eqn\nplusminusre{
\eqalign{n_+ &= 4 h_{21} + {2\chi\over 3} - 8N\cr
n_-^\phi &= 4h_{21} + \chi\cr}}
Notice that, as $\chi$ is divisible by 6 for manifolds of $SU(4)$
holonomy, $n_+$ is divisible by 4, which should be the case since 4
scalars fit into a supermultiplet. On the other hand, $\chi$ is not
divisible by 4 except under more stringent conditions such as elliptic
fibration\svw, so that $n_-^\phi$ is not necessarily divisible by 4.

Since, by virtue of supersymmetry, $n_+$ also counts the number of
$+$ chirality fermions, it only remains to count those of minus
chirality, which of course follows from the index theorem of the
previous section. The result is
\eqn\nminuspsi{
n_-^\psi = 4 h_{21}}

Observe that the IIA and IIB spectra written above are invariant under
mirror symmetry, which for eightfolds maps $h_{pq}$ to
$h_{4-p,q}$. This map preserves both $\chi$ and $h_{21}$, which are
the only invariants that determine the spectrum.

This completes the general analysis of the spectrum for type IIB on an
eightfold. We now describe a few examples, before turning to a
discussion of the relationship with type IIA after further
compactification on a circle.
\medskip

\noindent{\bf (i) Joyce Manifolds}

This is the case of $spin(7)$ holonomy, and $(0,2)$ spacetime
supersymmetry in 2d. The Joyce eightfolds\joyce\ are
blown-up orbifolds of the eight torus, $T^8/\Gamma$, where $\Gamma$ is
a suitable discrete group. It was shown by Joyce that there are
essentially five types of singularity in the space $T^8/\Gamma$ which
are to be blown up to construct the manifold.
 
The Joyce manifolds have $\chi=144$ and $\tau= 64$, verifying
Eq.\anomcanc\ with $N=2$. The spectrum is
\eqn\joycespec{
n_+ = 2+2b_2+b_4-64,\qquad 
n_-^{\phi} = 2+2b_2+b_4,\qquad
n_-^{\psi} = 2b_3}

\noindent{\bf (ii) Borcea eightfolds}

These are Calabi-Yau 4-folds of the form $(K3\times K3)/Z_2$ with
$SU(4)$ holonomy, and lead to $(0,4)$ supersymmetry for Type IIB
compactifications\ref\borcea{\BORCEA}. These manifolds are labelled
by a set of integers $(r_1, a_1, \delta_1; r_2, a_2,
\delta_2)$. The Euler characteristic and signature are
\eqn\borc{
\eqalign{
\chi &= 6(r_1-10)(r_2-10)+288\cr
\tau &= 2(r_1-10)(r_2-10)+128\cr}}
verifying Eq.\anomcanc\ with $N=4$. From the Hodge diamond one finds
\eqn\borchtwoone{
h_{21} = 5(r_1+r_2)-6(a_1+a_2)- \half(r_1 r_2 - a_1 a_2) + 22}
These data suffice to determine the spectrum. The mirror
transformation acts, for these manifolds, as $r_i\to 20 - r_i$ with
$a_i$ unchanged, and it is evident that the above data are invariant
under this.
\medskip

\noindent{\bf (iii) K3 $\times$ K3}

This is the case of $SU(2)\times SU(2)$ holonomy and $(0,8)$
supersymmetry. The relevant invariants are
\eqn\kthreesq{
\chi = 576, \qquad \tau = 256,\qquad h_{21} = 0}
verifying Eq.\anomcanc\ with $N=8$. Thus we have
\eqn\kthreesqspec{
n_+ = 320,\qquad n_-^{\phi} = 576,\qquad n_-^{\psi} = 0}
This reproduces the result of Ref.\ref\shibaji{\SHIBAJI}\ 
for $K3\times K3$.
\medskip

\noindent{\bf (iv) $T^8/Z_2$}

For this singular space, the holonomy is $Z_2$ and the supersymmetry
is $(0,16)$. The topological invariants have to be computed in the
orbifold sense, and one finds\ref\senorb{\SENORB}:
\eqn\teight{
\chi=384,\qquad \tau= 256,\qquad h_{21}=0} 
Thus the spectrum is
\eqn\teightspec{
n_+=128,\qquad n_-^{\phi}=384,\qquad n_-^{\psi}=0}
It has been noted in Ref.\ref\dm{\DM}\ that this compactification is
dual to the orientifold of M-theory on $T^9/Z_2$ (this and related
cases have been studied in more detail in Ref.\ref\alok{\ALOK}).
However, in the M-theory case one finds the spectrum in the form:
\eqn\mtnine{
n_+=128,\qquad n_-^{\phi}=128,\qquad n_-^{\psi}=512}
which is equivalent to the previous expression only after bosonization
of the 512 chiral fermions.

The different forms of Eqs.\teightspec\ and \mtnine\ are not
accidental, but closely related to the geometry of the corresponding
compactifying spaces. In the case of IIB on $T^8/Z_2$ we have
$2^8=256$ fixed points, and a chiral boson is obtained as the twisted
sector for each one. For M-theory on $T^9/Z_2$ there are instead
$2^9=512$ fixed points, and a chiral fermion arises as the twisted
sector for each one. Since M-theory does not possess a 1-brane, these
twisted sector multiplets must appear symmetrically at the fixed
points.

It is amusing that M-theory seems to ``know'' about bosonization. This
also has a nontrivial consequence which we will point out in the
following. 

\newsec{T-duality and the vacuum momentum}

Suppose we compactify type IIA and IIB on the same eightfold, and then
further on a circle to $0+1$ dimensions. T-dualizing along the circle
maps one theory to the other. Now we have an apparent puzzle: type IIA
has a 2-form tadpole in 2d, which will become a 1-form tadpole in 1d,
and this is proportional to the Euler characteristic $\chi$ of the
eightfold. However, we have found no inconsistency for type IIB on the
eightfold to two dimensions, so the inconsistency required by
T-duality must arise upon compactifying one further
dimension. Moreover, it must take the form of a tadpole for the KK
1-form $A=g_{12}$.

Recently it was pointed out\ref\ori{\ORI}, in the particular case of
$K3\times K3$ compactification, that this can be understood in terms
of the nonzero vacuum values of $L_0$ and ${\bar L}_0$ arising from
free fields on a cylinder. Here we will prove that the correct vacuum
momentum is obtained in the general case, and will examine this point
in a little more detail.

For 2d field theories on a cylinder, the generator of translations 
along the compact direction is $L_0 - {\bar L}_0$. Thus, a nonzero
value of this operator in the vacuum implies that, from a 2d point of
view, there is a nonzero momentum in the vacuum state. Under
T-duality, this will turn into a nonzero winding charge of the vacuum,
precisely what we would expect in a theory which has a 2-form tadpole
in 2d. The tadpole must have the precise value ${\chi\over 24}$.

Given the spectrum of type IIB in 2d, we use the fact that a free
periodic boson has vacuum energy $-{1\over 24}$ while a free periodic
fermion has energy $1\over 24$. Thus, associating $L_0$ to what we
have earlier called $+$ chirality, we have, for type IIB on an
eightfold, 
\eqn\lzero{
(L_0)_{\rm vac} = 0,\qquad ({\bar L}_0)_{\rm vac} = 
{1\over 24}(n_-^\psi - n_-^\phi)}
where the first equation is a consequence of the chiral supersymmetry
in 2d. From Eqs.\nplusminusre\ and \nminuspsi, it follows that
\eqn\lzeroval{
(L_0 - {\bar L}_0)_{\rm vac} = {\chi\over 24}}
as desired.

Note that if the circle becomes large and we are effectively in two
noncompact dimensions, this effect goes away. The reason is that the
operator $L_0$ as conventionally defined in conformal field theory has
a zero-point contribution $-{1\over 24}$ for a free boson only if the
radius of the circle (the range of the $\sigma$ coordinate) is fixed
to be $2\pi$, as is conventionally done. For a circle of radius $2\pi
R$, the zero-point contribution is actually $-{1\over 24 R}$, so that
it goes away in the limit $R\to\infty$. This explains why there is no
corresponding one-loop term in the effective action of type IIB theory
in 2 (or 6 or 10) dimensions, and yet the prediction of T-duality with
type IIA is satisfied.

\newsec{An M-theory subtlety}

The example of IIB on $T^8/Z_2$, and its M-theory dual discussed in
the previous section, now poses a slight problem. We saw that M-theory
produces the same spectrum but in a fermionized form. From Eq.\mtnine,
the vacuum momentum for M-theory on $T^9/Z_2$ compactified on a
further circle is actually $-{\chi\over 24}$, equal in magnitude but
opposite in sign to that of type IIB on $T^8/Z_2$. The sign can be
changed by redefining the coordinate $x^2$ to $-x^2$, but this
choice of convention is no longer available once we have fixed the
chirality of the spacetime supersymmetry. In other words, type IIB on
$T^8/Z_2$ produces a theory where the supersymmetry has $+$ chirality
and the vacuum momentum is positive, or else by a change of
convention, the supersymmetry has $-$ chirality and the vacuum
momentum is negative. On the other hand, M-theory on $T^9/Z_2$
gives a theory with supersymmetry of $+$ chirality and negative vacuum
momentum, or the other way around. This suggests that the equivalence
between the two theories is more subtle than was previously thought,
as long as the eightfold has nonzero $\chi$.

This may seem surprising given that the two theories are related by
bosonization, but the reason is that the fermion contribution to the
zero-point values for $L_0$ and ${\bar L}_0$ has been calculated for
periodic boundary conditions on the fermions around the circle. But bosonization does not
equate periodic bosons with periodic fermions.

One may expect a larger class of dualities between IIB on orbifolds of
the form ${\cal M}_8/Z_2$ and M-theory on corresponding orientifolds
$({\cal M}_8\times S^1)/Z_2$. In these cases, the latter theory will
always have twice the number of fixed points as the former, since the
extra $S^1$ contributes a pair of fixed points for each one of the
original orbifold. Assuming a symmetric distribution of twisted-sector
states, it would seem that M-theory should always give the fermionized
form of the IIB compactification, as in the case we have
discussed. This would be interesting to confirm.
\bigskip

\noindent{\bf Acknowledgements}

\noindent We are grateful to Ashoke Sen and especially Atish Dabholkar
for helpful discussions. We acknowledge useful correspondence with
K. Becker, M. Becker, O. Ganor and E. Witten.

\listrefs
\end